\documentclass[twocolumn,eqsecnum,prb]{revtex4}
\usepackage{amsfonts}
\usepackage[T1]{fontenc}
\usepackage{amsmath,amsbsy,amssymb,graphicx,braket}
\usepackage{times}

\def\lamSO{\lambda_{\text{SO}}}
\def\lamV{\lambda_V}

\begin{document}

\title{{\Large Edge States in Silicene Nanodisks}}
\author{Ko Kikutake$^1$, Motohiko Ezawa$^1$ and Naoto Nagaosa$^{1,2}$}
\affiliation{$^1$Department of Applied Physics, University of Tokyo, Hongo 7-3-1, 113-8656, Japan}
\affiliation{$^2$Center for Emergent Matter Science (CEMS), \\
ASI, RIKEN, Wako 351-0198, Japan}

\begin{abstract}
	Silicene is a honeycomb-structure silicon atoms, which shares many intriguing properties with graphene. Silicene is expected to be a quantum spin-Hall insulator due to its spin-orbit interactions.
	We investigate the electronic properties of silicene nanodisks, which are silicene derivatives with closed edge.
	In case of the simplest model of graphene nanodisks, the number of the zero-energy modes is given by the lower bound of the Lieb theorem.
	They are standing wave states.
	When the spin-orbit interaction is introduced and the system becomes a topological insulator, they begin to propagate around a nanodisk. 
	Helical edge currents flow around the edge of a nanodisk though the crystal momentum is ill-defined.
	We have newly found the counting rule of the zero-energy states, and constructed the low-energy theory of zigzag triangular silicene.
	We also show the validity of the bulk-edge correspondence in nanodisks with rough edge by calculating the local probability amplitude and current. 
	Finally, the signatures of the topological phase transition in silicene nanodisks under an external electric field are revealed.
	Our results will be observable by means of STM experiments.
\end{abstract}

\maketitle
\section{Introduction}
	Topological insulator is one of the most fascinating concepts in physics. 
	It is characterized by the bulk-edge correspondence: 
	Gapless edge modes emerge necessarily along an edge of a topological insulator.
	The edge states of the nanoribbon geometry are well studied. 
	However there are few reports on the bulk-edge correspondence on a nano-scale object with a closed edge, that is a nanodisk, especially with a rough edge. 
	Strictly speaking the topological number is ill-defined for nanodisks since it is given by the integral over the Berry connection on the whole Brillouin zone, but it is absent. 
	Nevertheless, when a nanodisk is sufficiently large, it must have all properties of a topological insulator.
	Thus a natural question is the size effects on the topological properties.
	We investigate this point by studying nanodisks made of silicene.
	
	Silicene is a honeycomb structure made of silicon atoms. 
	The low-energy dynamics is described by the Dirac theory as in graphene. However, Dirac electrons are massive due to a relatively large spin-orbit gap in silicene.
	It attracts much attention since silicene is expected to be topological insulators such as quantum spin-Hall states\cite{Yao201102} and quantum anomalous Hall states\cite{Ezawa20122,Ezawa2013L,Ezawa2013B} depending on externally controlable parameters. 
	Indeed, the band gap can be tuned by applying electric field\cite{Ezawa20121}.

	Silicene is a topological insulator because of the spin-orbit interaction (SOI).
	As far as the Hamiltonian concerns, the dynamical properties of silicene are obtained from those of graphene by introducing the SOI and the Rashba interaction (RI)\cite{Yao201101}.
	Graphene nanoribbons show salient electronic properties depending on the edge structure (zigzag or armchair) and the width\cite{Nakada,Fertig,Ezawa20061}. 
	Flat bands localized at edge emerge in zigzag graphene nanoribbons. 
	Silicene nanoribbon presents an interesting playground to study the bulk-edge correspondence\cite{EzawaNagaosa}.
	With respect to graphene nanodisk, there appear zero-energy edge  states\cite{rossier,Ezawa20071,Wang,Yazyev2010}, whose number $\xi_0$ is given by
$ \xi_0 = 2|N_A-N_B|$ with $N_A$ and $N_B$ being the number of sites in $A$ and $B$ sublattices, respectively, in accord with the Lieb theorem\cite{Lieb} valid on the bipartite lattice.
	Let us call them the Lieb modes.
	Our concern is the fate of these Lieb modes when SOI and then RI additionally are introduced into the Hamiltonian in view of the bulk-edge correspondence.


	In this paper, we  investigate how the energy spectrum of a nanodisk with zigzag, armchair or rough edge evolves as SOI changes. Because of the electron-hole and time-reversal symmetry there is at least one Kramers pair of zero-energy states when the number of the lattice sites is odd. By calculating the local density of state, we show the wave function is localized along the edge. We also show that up and down spin flows counter-clockwisely even though we can not define the crystal momentum in nanodisk. 	The emergence of the helical current is a manifestation that silicene is a QSH insulator. This is in contrast to the fact that  the edge states of graphene nanodisks are standing wave.

\section{Hamiltonian}
	We investigate the tight-binding model described by
\begin{align}
		H &= -t\sum_{\langle i,j\rangle s} c^{\dagger}_{is}c_{js}
		+i\frac{\lambda_{\mathrm{SO}}}{3\sqrt{3}}\sum_{\langle\langle 
i,j\rangle\rangle ss'} \nu_{ij}c^{\dagger}_{is} \sigma^z_{ss'}c_{js'}\notag\\
		&-i\frac{2}{3}\lambda_{\mathrm{R}} \sum_{\langle\langle 
		i,j\rangle\rangle ss'} \nu_z^i c^{\dagger}_{is} (\mathbf{\sigma}\times\hat{\mathbf{d}}_{ij})_{ss'}^zc_{js'}
		+ \lambda_V \sum_{is} \nu_z^i c^{\dagger}_{is}c_{is},\label{eq:hamiltonian}
\end{align}
	where $c^{\dagger}_{is}$ is the creation operator of an electron at $i$ site with spin $s$, and $\langle i,j\rangle/ \langle\langle
	i,j \rangle\rangle$ runs over all pair of sites which is nearest-neighbor/next-nearest-neighbor, and $\mathrm{\sigma}$ denote the 
	Pauli matrices.
	The first term is the nearest-neighbor spin-independent hopping (NNH) with transfer energy $t$.
	The second term represents the SOI with coupling parameter $\lambda_{\text{SO}}$, 
	and $\nu_{ij}=+1$ if the next-nearset-hopping is counterclockwise around a hexagon, otherwise $\nu_{ij}=-1$.
	The third term expresses the RI with coupling parameter $\lambda_{R}$, 
	$\nu_z^i = \pm 1$ for $i$ representing the A or B site, and 
	$\hat{\mathbf{d}}_{ij} = \mathbf{d}_{ij}/|\mathbf{d}_{ij}|$ with $\mathbf{d}_{ij}$ the vector from the $j$ site to the $i$ site.
	The last term is the staggered potential with coupling $\lambda_V$.
	We denote each term as $H_{\mathrm{NNH}}$, $H_{\mathrm{SOI}}$, $H_{\mathrm{RI}}$, $H_{\mathrm{V}}$ for
	NNH term, SOI term, RI term, staggered term, respectively.

	
	For the bulk system, the Hamiltonian describes a topological insulator when 
	$0<|\lambda_{V}|<\lambda_{\mathrm{SO}}$, 
	and a trivial insulator phase when 
	$\lambda_{\mathrm{SO}}<|\lambda_V| $.
	We note that graphene is described by the Hamiltonian (\ref{eq:hamiltonian}) with 
	$\lambda_{\mathrm{R}}=\lambda_{\mathrm{SO}}=\lambda_V=0$.
	
	The particle-hole symmetry operator is defined by
	\begin{align}
	\Xi c_{As}\Xi ^{\dag }=c_{As}^{\dag },\quad \Xi c_{Bs}\Xi ^{\dag }=-c_{Bs}^{\dag},
	\end{align}
	Under the transformation, the Hamiltonian  $H_{\mathrm{NNH}} + H_{\mathrm{SOI}}$ is invariant, but $H_{\mathrm{NNH}}+H_{\mathrm{SOI}} + H_{\mathrm{RI}}$ is not. Hence, as far as only the SOI is introduced, the system has the particle-hole symmetry. However, if the RI is additionally introduced, the symmetry is broken. The deviation from the zero-energy states due to the RI is almost negligible since the effect of RI vainishes at K and K' points. Also $\lambda_V$ breaks the particle-hole symmetry.
	On the other hand, the time-reversal symmetry exists in the Hamiltonian (\ref{eq:hamiltonian}).

\begin{figure}[t]
\centerline{\includegraphics[width=0.49\textwidth]{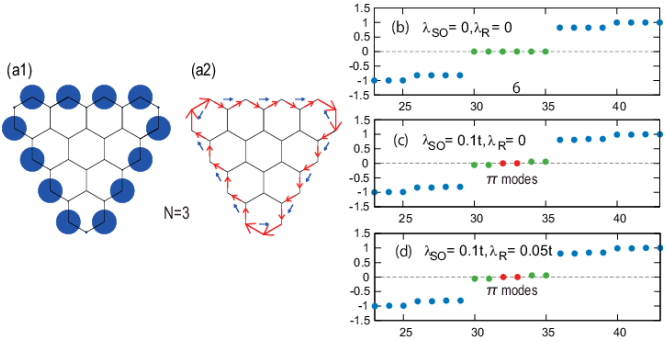}}
\caption{(a1),(a2) Local probability amplitude and current of the up-spin $\pi$ mode. We have taken $\lambda_{\mathrm{SO}}=0.1t,\lambda_{\mathrm{R}}=0$ for illustration.
(b),(c),(d) Energy spectra with parameters shown within figures. 
The vertical axis is $E/t$, and the horizontal axis is the index of the eigenstates. 
We have taken $N_A=18,N_B=15$. There are $6$ Lieb modes (green) in the graphene nanodisk and $2$ $\pi$ modes (red) in the silicene nanodisk.}
\label{tri_4}
\end{figure}

	We define the number of positive, zero and negative energy eigenstates as 
	$\xi_p$, $\xi_n$ and $\xi_0$, respectively.
	When the particle-hole symmetry exists in the system, the numbers of the positive and negative energy are equal $\xi_p=\xi_n$. The number of the zero energy states is determined as $\xi_0=N-\xi_p-\xi_n=N-2\xi_p$ in the spinless theory. If the total site number $N\equiv N_A+N_B$ is odd, there is at least one zero-energy state. 
	In the presence of the spin degree of freedom, the states appear as in pairs, which are known as the  Kramers pair. Then, $\xi_p$ and $\xi_0$ are even.
	This counting rule tells that there exists at least one zero-energy Kramers pair when $N$ is odd.
	Consequently, we obtain the following counting rule on the number of zero-energy states,
	\begin{equation}
	\xi_0 \geq 2 \times (|N_A-N_B|\,\mathrm{mod}\,2),	\label{Count}
	\end{equation}
	instead of the Lieb theorem. 
	Actually we find the lower bound of this counting rule to hold as far as we have checked numerically.   
	We refer these two zero modes as the $\pi$ modes for odd $N$, as illustrated in Fig.\ref{tri_4}, and also the two modes nearest to the Fermi level in the valence band as the $\pi$ modes for even $N$, as illustrated in Fig.\ref{tri_5}.
	We can present an intuitive picture of this formula for zigzag triangular nanodisks, as we shortly show: See Fig.\ref{FigGuki}.

\section{Zigzag triangular silicene} 
\begin{figure}[t]
\centerline{\includegraphics[width=0.49\textwidth]{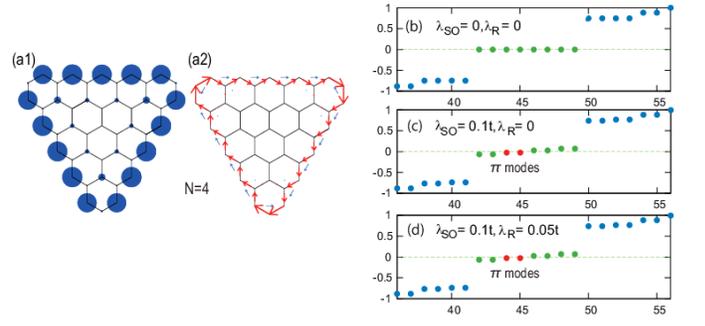}}
\caption{See the caption of Fig.\ref{tri_4}. Here we have taken $N_A=25,N_B=21$.}
\label{tri_5}
\end{figure}

First we investigate zigzag triangular silicene.
	We show the energy spectrum in Fig.\ref{tri_4} for the case of $N_A=18,N_B=15$. 
	The zigzag triangular graphene has 6 zero-energy modes, in accord with the Lieb theorem.
	When we include the SOI, there remains two zero modes ($\pi$ modes) with the other four modes acquiring nonzero energies proportional to $\lambda_{\text{SO}}$. 
	When we include the RI additionally, these two zero-energy modes disappear. 
	Nevertheless, they remain very near to the zero energy.
	We also show the energy spectrum in Fig.\ref{tri_5} in the case of $N_A=25,N_B=21$. 
	Although there exists no zero-energy modes, we can identify the two modes ($\pi$ modes) nearest to the Fermi level from the valence band.
	The $\pi$ modes can be identified with the gapless edge modes in a sufficiently large nanodisk.

	Next, we investigate the wavefunctions of the $\pi$ modes and how the bulk-edge correspondence is realized by them.
	Fig.\ref{tri_4} and Fig.\ref{tri_5} represent the energy spectrum of a triangular graphene and silicene for $N_A=18,N_B=15$ and $N_A=25,N_B=21$, respectively.
	We also show the local probability amplitude $|\psi_i|$ 
	and the local probability current 
	\begin{equation}
	J_{ij}=2\mathrm{Im}\psi^*_iH_{ij}\psi_j 
	\end{equation}
of the $\pi$ mode state with $\lambda_{\mathrm{SO}}$.
	 They satisfy the continuous equation 
	\begin{equation}
	 \frac{\mathrm{d}\rho_i}{\mathrm{d}t} + \sum_j J_{ij}=0 
	\end{equation}
	 with $\rho_i=\psi_i^*\psi_i$.
	 The penetration depth of zigzag edge is as short as the atomic scale in the case of zigzag edges, which is consistent with the result of silicene nanoribbons\cite{EzawaNagaosa}.
	
	The $\pi$ mode is localized along the edge.
	The edge states of graphene nanodisk are standing waves ($J_{ij}=0$). On the other hand, once we include the SOI, the current begins to flow ($J_{ij}\not=0$). 
	The resultant flow is helical due to the time-reversal symmetry.
	The local probability current of the up-spin $\pi$ mode circulates along the edge in the clockwise direction, and that of the down-spin $\pi$ mode in the anticlockwise direction.
	The emergence of the helical current is a manifestation that silicene is a QSH insulator.

\begin{figure}[t]
\centerline{\includegraphics[width=0.4\textwidth]{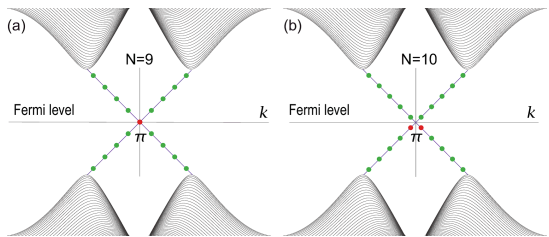}}
\caption{Illustration of the correspondence between the edge modes of a zigzag nanoribbon and a zigzag triangular nanodisk for $N=9$ and $N=10$. The edge modes represent almost straight lines connecting the tips of the Dirac cones in the conduction and valence bands in a nanoribbon. The green dots represent the energy spectrum of a nanodisk. We focus on one edge of a nanodisk. Since the length is finite, the momentum is quantized. When $N$ is odd, there exist two states at the $\pi$ point on the Fermi level, which we call the $\pi$ modes. When $N$ is even, there exist four states near the $\pi$ point in the valence bands, which we also call the $\pi$ modes.
}
\label{FigGuki}
\end{figure}
	
	There is an intuitive and intriguing correspondence between the energy spectrum of a nanodisk and a nanoribbon\cite{Ezawa20101}, as illustrated in Fig.\ref{FigGuki}.
	The zero-energy states of a zigzag triangular graphene nanodisk are constructed by quantizing the energy spectrum of the edge states of a zigzag graphene nanoribbon. 
	By imposing that the wave functions vanishe at the corners of the triangular nanodisk, the momentum is quantized as 
	\begin{equation}
	ak_n=\pm[(2n+1)/3N+2/3]\pi
	\end{equation}
	with $0\leq n\leq(N-1)/2$. 
	A zigzag graphene nanoribbon has flat bands in $2\pi/3<|ak|<\pi$. 
	When $N$ is even, there are $N/2$ states for $k_n>0$ and $N/2$
states $k_n<0$. When $N$ is odd, there are $(N-1) /2$ states for
$k_n>0$ and $(N-1) /2$ states for $k_n<0$. Additionally, there exists $\pi$ modes for $n=(N-1)/2
$ when $N$ is odd.
	Hence we conclude that the number of the zero-energy states is $N$ without including spin degeneracy. They are the Lieb modes.
	They are standing waves corresponding to those of a graphene nanoribbon. 
	The standing waves become helical modes as the SOI is introduced.
	We may still require the same momentum quantization, and find that there exists two states with $ak=\pi$ on the Fermi level when $N$ is odd [Fig.\ref{FigGuki}(a)], while there are none when $N$ is even but there are two nearest to the Fermi level [Fig.\ref{FigGuki}(b)]. 
	We have called them the $\pi$ modes owing to the condition $ak=\pi$.

\section{Triangular silicene with a defect}
\begin{figure}[t]
\centerline{\includegraphics[width=0.49\textwidth]{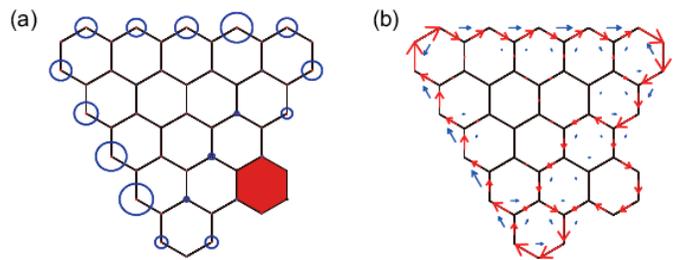}}
\caption{(a) Local density of the $\pi$ mode in an triangular silicene with a defect. 
The blue circle represents the magnitude of the local density. The defect of the edge are colored in red. 
(b) Local current of the $\pi$ mode.
Red (blue) arrows indicate flows between the nearest-neighbor (second-nearest-neighbor) sites, respectively. We have taken $\lambda_{\text{SO}}=0.1t$ for illustration.}
\label{tri_rough_5_1}
\end{figure}

	We show the energy spectra for a zigzag triangle with a defect in Figs.\ref{tri_rough_5_1}. 
	In the figure, we can see that a flow separates into two streams.
	One stream flows along an edge of the defect.
	Meanwhile, the other stream flows in the inside of the triangle.
	After the defect, two streams merge into one stream. This can be understood as follows. The defect locally produce armchair edges. In the armchair region, the penetration depth is much larger than that of zigzag edges. Thus the wave function penetrates into deeply near the defect.
	
\section{Nanodisks with rough edge}
	We proceed  to study the bulk-edge correspondence in a general silicene nanodisk with a rough edge. A rough edge consists of zigzag edge parts and armchari edge parts.
	Even in this case, for graphene we have the Lieb modes, and for silicene
	we can identify the two $\pi$ modes either on the Fermi level or nearest to the Fermi level in the valence band.
	
\begin{figure}[t]
\centerline{\includegraphics[width=0.49\textwidth]{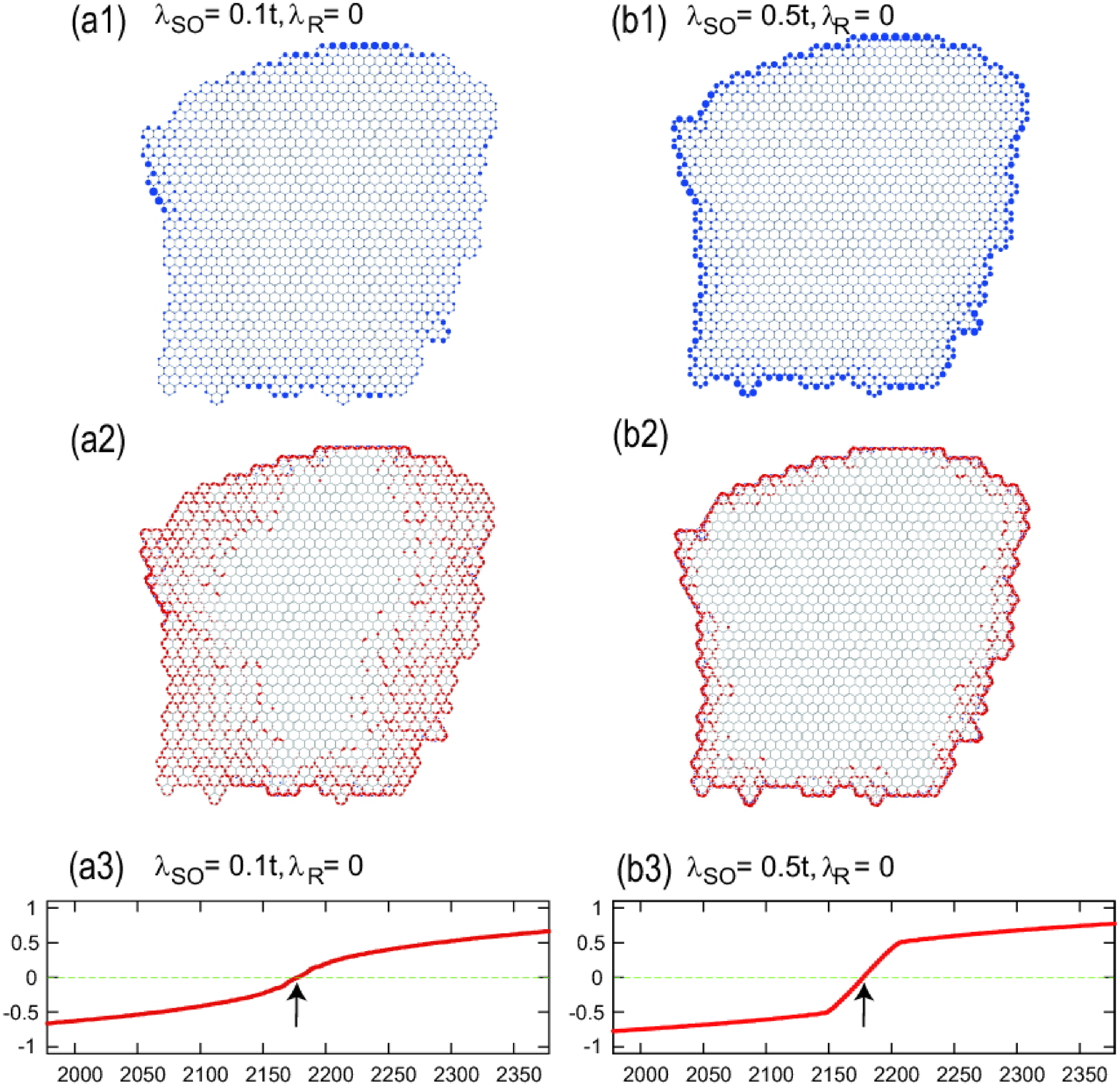}}
\caption{Local probability amplitude (a1), (b1) and local current (a2),(b2) of the up-spin $\pi$ mode.
The local probability current circulates in the clockwise direction.
(a3),(b3) Energy spectra.
The vertical axis is $E/t$, and the horizontal axis is the index of the eigenstates. 
We have taken $N_A=1087$ and $N_B=1091$. There exist $8$ Lieb modes.}
\label{tri_rough_big}
\end{figure}

	First of all, there is no local probability current in graphene, showing that only standing wave states are present. 
	When $\lambda_{\text{SO}}$ is introduced, the standing wave begin to propagate.
	This is again a manifestation of the bulk-edge correspondence.
	In Fig.\ref{tri_rough_big}, we show the local probability amplitude and the local probability current of the $\pi$ mode with several values of $\lambda_{\text{SO}}$.
	When $\lambda_{SO}=0$, the Lieb modes are localized along the zigzag edge. As $\lambda_{SO}$ is introduced, they begin to propagate and penetrate into the regions of armchair edges, while they are rather loosely bound perpendicular to the armchair edges as shown in Fig.5(a1). The zero energy modes are two of these Lieb modes.
	A rough edge is divided into zigzag and armchair edges locally. 
	In the local zigzag region, the penetration is as short as the lattice constant, 
	while in the local armchair region, the penetration length is antiproportional to the magnitude of the SOI\cite{EzawaNagaosa}, 	
	\begin{equation}
	L_{\text{arm}}=\hbar v_{\text{F}}/\lambda_{\text{SO}}, 
	\end{equation}
	which is much larger than that of zigzag edges.	
	The local current circulates around a nanodisk, as seen in  Fig.\ref{tri_rough_big}(a2).
	However, the local densities are separeted and the local probability current penetrates into bulk.
	By increasing the SOI, the spatially localized DOS is elongated along the edge.
	When it is sufficiently large, the $\pi$ mode is localized along the edge, and the local probability current flows along the edge [Fig.\ref{tri_rough_big}(b1,b2)].

\section{Triangular silicene with staggered potential}

\begin{figure}[t]
\centerline{\includegraphics[width=0.5\textwidth]{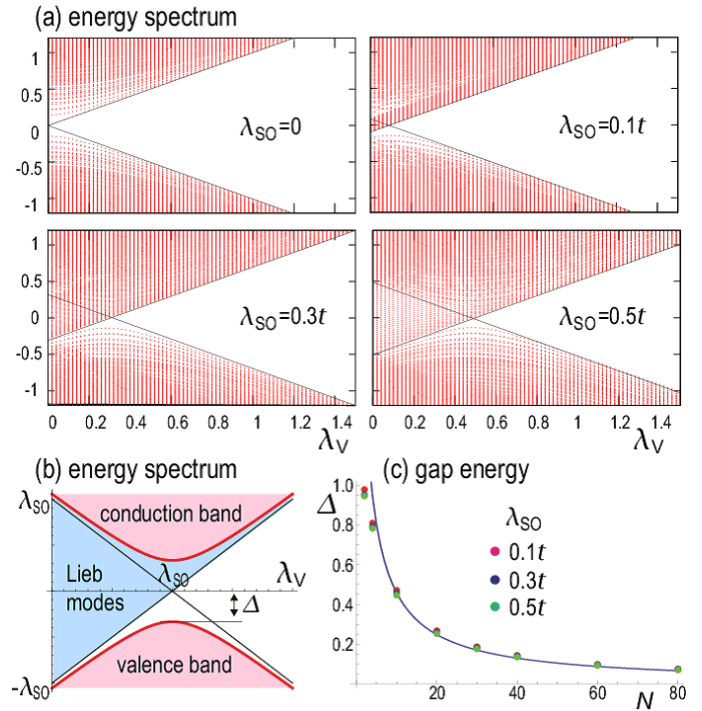}}
\caption{(a) Energy spectrum as a function of $\lambda_V$ for the system with $\lambda_{\text{SO}}$ as indicated. The system size is $N=40$. The vertical axis is the energy in unit of $t$.
The two lines are linear extrapolations from the band spectrum, which cross each other at the phase transition point of the bulk system. 
(b) Schematic illustration of the energy spectrum.
The energy levels between  the two lines for $\lambda_V<\lambda_{\text{SO}}$ represents the Lieb modes. The red curve is the extremum energy $\pm E(\lamV)$ in the energy spectrum.
given by (\ref{sqrt}), while the thin lines are $E=\pm(\lamV - \lamSO)$.
(c) Numerical fitting of $\Delta$ by the curve given by (\ref{DeltaN}).
}
\label{sta}
\end{figure}

	Finaly, we investigate the effects of the staggered potential induced by an external electric field, which breaks the particle-hole symmetry.
	Silicene undergoes a topological phase transition from a QSH to a trivial insulator by applying electric field\cite{Ezawa20121}. 
	It is important to study how this phase transition is observed in silicene nanodisks.

In Fig.\ref{sta}(a), we show the energy spectra with $\lamSO=0, 0.1t, 0.3t, 0.5t$
for zigzag triangular silicene.
We explain typical features.
First, the minimum (maximum) of the spectrum in the conduction (valence) band is well fitted 
by the stright line, $E=\pm(\lamV - \lamSO)$. 
They meet the topological phase transition point $\lamV=\lamSO$. 
Second, there are many states inside the bulk gap for $\lamV<\lamSO$.
They are the Lieb modes, which turn into edge channels, as illustrated shematically in Fig.\ref{sta}(b). 
The energy of Lieb modes changes linearly as a function of staggered potential. This is due to the fact that the Lieb modes are localized only at one sublattice.

The actual bulk band gap does not close due to finite size effects.
We may say that the topological phase transition is rounded in nanodisk geometry.
However, it must manifest itself clearly for sufficiently large nanodisks.
We wish to determine the crossover size beyond which the signal of the topological phase transtion is manifest.

	To investigate the finite size effect of topological phase transition quantitatively, 
we focus on the maximum energy $E(\lamV)$ in the valence band as a function of the staggerd potential.
In Appendix we propose to fit it as	
\begin{equation}
E(\lamV) = \sqrt{\alpha t^2/N_{\text{tot}}+(\lambda_{\text{SO}}-\lambda_{V})^2},
\label{sqrt}	
\end{equation}
where $\alpha$ is a phenomenological parameter. 
We set a gap due to the finite size effects,
\begin{equation}
\Delta=|E(\lamSO)|=\sqrt{\frac{\alpha t^2}{N_{\text{tot}}}}
=\sqrt{\frac{\alpha t^2}{N^2+4N+1}},
\label{DeltaN}
\end{equation}
as illustrated in Fig.\ref{sta}(b). This is one half of the band gap.

We have numerically fitted $\Delta$ as a function of $N$ for various $\lamSO$.
The fitting is very good by choosing $\alpha =29.6$ for all values of $\lamSO$ as far as we have checked.
When $\Delta >\lamSO$, the finite size effects are dominant than the SOI. On the other hand, when $\Delta <\lamSO$, we can observe a signature of the topological phase transition. Namely, there is a crossover around $\Delta =\lamSO$ whether we can observe topological phase transition.
We can estimate the size of this crossover. 
The condition on the crossover size would be given by $\Delta=\lamSO$, from which we find
\begin{equation}
N_{\mathrm{tot}}=\alpha t^2/\lambda_{\mathrm{SO}}^2.
\end{equation}
The crossover size is $N=9$ for $\lambda_{\text{SO}}=0.5t$ and $N=52$ for $\lambda_{\text{SO}}=0.1t$.
The real value corresponds to $\lambda_{\text{SO}}=0.002t$. The corresponding size is $N=2718$, which is approximately $1\mu m$.

	In conclusion, we have investigated edge states in silicene nanodisks with the help of the particle-hole symmetry and the time-reversal symmetry.
	We have newly found the counting rule of the zero-energy states, where the existence of the zero-energy states is protected by the particle-hole symmetry when the lattice number is odd and the RI is zero. 
	We have constructed the low-energy theory of zigzag triangular silicene.
	Helical edge currents flow around the edge of a nanodisk though the crystal momentum is ill-defined.
	Next we have studied nanodisks with rough edge. 
	We have found the validity of the bulk-edge correspondence in silicene nanodisks in general.
	The local density of a silicene nanodisk with rough edge can be interpreted by the difference of the penetration length of zigzag and armchair edges. Finally, we found the topological phase transition becomes crossover in silicene nanodisks due to finite size effects. We have determined the crossover size as a function of $\lambda_{\text{SO}}$.

\appendix
\section{Effective theory}
	In the following we estimate the extremum energy $E(\lamV)$ in the band spectrum.
In the Dirac theory the density of states is given by 
\begin{equation}
\rho (E) =\frac{4|E|}{\sqrt{3}\pi t^2}\theta (|E|-|\lambda_{\text{SO}}-\lambda_{V}|)
\end{equation}
with the step funtion $\theta (x)=1$ for $x>0$ and $\theta (x)=0$ for $x<0$.
The number of the states $n$ between the energy $E$ and the Fermi energy is given by
\begin{align}
n(E)&=N_{\text{tot}}\int_0^E\rho(\varepsilon )d\epsilon \\
&=\frac{2N_{\text{tot}}}{\sqrt{3}\pi t^2}(E^2-(\lambda_{\text{SO}}-\lambda_{V})^2)\theta (|E|-|\lambda_{\text{SO}}-\lambda_{V}|),
\end{align}
where the number of the total sites is $N_{\text{tot}}=N^2+4N+1$.
The inversion formula reads
\begin{equation}
E(n)=\pm\sqrt{\Delta^2 +(\lambda_{\text{SO}}-\lambda_{V})^2},
\end{equation}
where $\Delta$ is given by (\ref{DeltaN}) with $\alpha=\sqrt{3}\pi nt^2/2$. 
Strictly speaking the above analysis is valid for sufficiently large nanodisks.
Hence it is not justified to apply this to determine the extremum energy by choosing $n=1$.
We propose to use this as a phenomenological formula with $\alpha$ a phenomenological parameter.


\begin{thebibliography}{99}
	\bibitem{KaneMele} C. L. Kane, and E. J. Mele, Phys. Rev. Lett. \textbf{95},	226801 (2005).
	\bibitem{Lay} P. Vogt, P. D. Padova, C. Quaresima, J. Avila, E. Frantzeskakis, M. C. Asensio, A. Resta, B. Ealet, and G. L. Lay, Phys. Rev. Lett. \textbf{108}, 155501 (2012).
	\bibitem{Takamura} A. Fleurence, R. Friedlein, T. Ozaki, H. Kawai, Y. Wang, and Y. Yamada-Takamura, Phys. Rev. Lett. \textbf{108}, 245501 (2012).
	\bibitem{Kawai} C.-L. Lin, R. Arafune, K. Kawahara, N. Tsukahara, E. Minamitani, Y. Kim, N. Takagi, and M. Kawai, Appl. Phys. Express \textbf{5}, 045802 (2012).
	\bibitem{Wu1} B. Feng, Z. Ding, S. Meng, Y. Yao, X. He, P. Cheng, L. Chen, and K. Wu, Nano Lett. \textbf{12}, 3507 (2012).
	\bibitem{Yao201102} C.-C. Liu, W. Fenng and Y. Yao, Phys. Rev. Lett. \textbf{107}, 076802 (2011).
	\bibitem{Ezawa20122} M. Ezawa, Phys. Rev. Lett. \textbf{109}, 055502 (2012).	
	\bibitem{Ezawa2013L} M. Ezawa, Phys. Rev. Lett. \textbf{110}, 026603 (2013).
	\bibitem{Ezawa2013B} M. Ezawa, Phys. Rev. B \textbf{87}, 155415 (2013).
	\bibitem{Ezawa20121} M. Ezawa, New J. Phys. \textbf{14}, 033003 (2012).
	\bibitem{Yao201101} C.-C. Liu, H. Jiang, Y. Yao, Phys. Rev. B. \textbf{84}, 195430 (2011).
	\bibitem{Nakada} K. Nakada, M. Fujita, G. Dresselhaus and S. Dresselhaus, Phys. Rev. B \textbf{54} 17954 (1996).
	\bibitem{Ezawa20061}M. Ezawa, Phys. Rev. B \textbf{73}, 045432 (2006).
	\bibitem{Fertig} L. Brey and H. A. Fertig, Phys. Rev. B, \textbf{73}, 235411 (2006).
	\bibitem{Son} Y.-W. Son, Marvin, L. Cohen and S. G. Louie, Nature \textbf{444}, 347 (2006).
	\bibitem{Ezawa20071} M. Ezawa, Phys. Rev. B \textbf{76}, 245415 (2007).
	\bibitem{rossier} J. Fern\'andez-Rossier and J. J. Palacios, Phys. Rev. Lett. \textbf{99} 177204 (2007).
	\bibitem{Wang} W. L. Wang, O. V. Yazyev, S. Meng, E. Kaxiras, Phys. Rev. Lett. \textbf{102} 157201 (2009).
	\bibitem{Yazyev2010} O. V. Yazyev, Rep. Prog. Phys. \textbf{73} 056501 (2010).
	\bibitem{Lieb} E. H. Lieb, Phys. Rev. Lett. \textbf{62}, 1201 (1989).
	\bibitem{EzawaNagaosa} M. Ezawa and N. Nagaosa, cond-mat/arXiv:1301.6337.
	\bibitem{Ezawa20101} M. Ezawa, Phys. Rev. B \textbf{81}, 201402(R) (2010).
	\bibitem{Sachs} S. Fajtlowicz, P. E. John, and H. Sachs, Croat. Chem. Acta \textbf{78}, 2 (2005).
	\bibitem{Kaxiras2009} W. L. Wang, O. V. Yazyev, S. Meng, E. Kaxiras, Phys. Rev. Lett. \textbf{102} 157201 (2009).
\end{thebibliography}
\end{document}